\documentclass[12pt]{iopart}
\usepackage{graphicx}
\usepackage{iopams}
\usepackage{setstack}
\usepackage{bm}
\input{epsf}

\def\beq{\begin{equation}}
\def\eeq{\end{equation}}
\def\ba{\begin{eqnarray}}
\def\ea{\end{eqnarray}}

\def\b0{\arrowvert_{\beta=0}}

\begin{document}
\title{Fractal diffusion coefficient from dynamical zeta functions} 
\vskip4mm

\author{Giampaolo Cristadoro}
\address{Max Planck Institute for the Physics of Complex Systems, N\"{o}thnitzer Str. 38, D 01187 Dresden, Germany}
\eads{\mailto{giampo@mpipks-dresden.mpg.de}}

\begin{abstract}
Dynamical zeta functions provide a powerful method to analyze low dimensional dynamical systems when the underlying symbolic dynamics is under control.  On the other hand even simple one dimensional maps can  show an intricate structure of the grammar rules that may lead to a non smooth dependence  of global observable on parameters changes. A paradigmatic example is the fractal diffusion coefficient arising in a simple piecewise linear one dimensional map of the real line. Using the Baladi-Ruelle generalization of the Milnor-Thurnston kneading determinant we provide the exact  dynamical  zeta function for such a map and compute the diffusion  coefficient from its smallest zero.
\end{abstract}

\maketitle
Chaotic dynamic forces to move the attention from individual trajectory to evolution of densities. The study of the spectrum of the evolution operator, the Perron- Frobenius operator, is a well established method for the analytic treatment of hyperbolic dynamical systems. In particular dynamical averages can be extracted from the spectra of the generalized transfer operator: this approach was  successfully used to compute physical quantities like i.e. the diffusion coefficient for extended hyperbolic dynamical systems or escape rate for open systems \cite{POE}.  Relaxing the assumption of full hyperbolicity is  the natural step forward: lot of efforts are still devoted to understand how to extend the assumptions of strong chaoticity to include the presence of regions of regular motion into a meaningful periodic orbit expansion.  Along this line it was possible to show how to treat systems  with marginal periodic orbits \cite{inter}. In such a case marginality have a deep effect on the analytical structure of zeta functions and the system shows peculiar behavior  like anomalous diffusion and power law decay of correlations \cite{vari}. 

Even without relaxing the hyperbolicity assumption a strong limitation on the practical implementation of the periodic orbits program relies on the need of control over periodic orbits proliferation: the number of periodic orbits grows exponentially with the period (at a rate given by the topological entropy) and  an analytical treatment  requires  the control over  such a huge amount of information. Cycle expansions technique prevents the explosion of the theory organizing cycles into fundamental terms and curvature corrections (constructed with longer cycles and their shadowing by shorter orbits)  that comes out to be a good perturbative expansion \cite{POE}. The main ingredient is then a good understanding of the cycles that build the fundamental  and curvature terms  that implies a good control over the  symbolic dynamic of the system. Unfortunately the missing of such a control is the normal situation rather than the exception. Even for a very simple one dimensional map without complete branches such a control is typically impossible. As a paradigmatic example  we  will use a simple one dimensional linear map on the real line where it was shown \cite{Klages-Dorfman} that the diffusion coefficient is a fractal function of the parameter defining it. Such a complex behaviour is the result of the intricacy of the symbolic dynamics for a generic parameter value. Even if transport properties of such maps were already derived in \cite{Gro} any attempt to analytical organize the cycle expansions for the full parameter range has failed so far: at our knowledge until now it was only possible to identify a family of parameter values were symbolic dynamics become tractable and dynamical zeta function can be derived \cite{japan}. Here we show how it is possible to derive the diffusion coefficient for the full range of parameter values  using an alternative formulation of the dynamical zeta functions presented by Baladi-Ruelle \cite{BaRu} (their approach generalize the Milnor-Thurnston work \cite{MiThu} originally devoted to topological zeta function only). 

\section{Invariant coordinates and kneading determinants}
For $\Lambda \ge 2$ let's define the map $g_{\Lambda}:\mathbb{R}\to \mathbb{R}$ as a lift of the map $\hat{g}_{\Lambda}:[0,1]\to \mathbb{R}$:
\begin{eqnarray}
&g_{\Lambda}(x+n)=\hat{g}_{\Lambda}(x)+n \qquad n \in \mathbb{Z} \\
&\hat{g}_{\Lambda}(x)=\frac12+\Lambda (x-\frac12)
\end{eqnarray}
Let's define $f_{\Lambda}(x): [0,1]\to [0,1]$ to be the restriction on the unit interval of $\hat{g}_{\Lambda}(x)$ i.e.
\begin{equation}
 f_{\Lambda}(x)=\hat{g}(x) \quad {\rm{mod}} \, 1
 \end{equation}

\begin{figure}[h!]  
\centerline{\epsfxsize=9cm\epsfbox{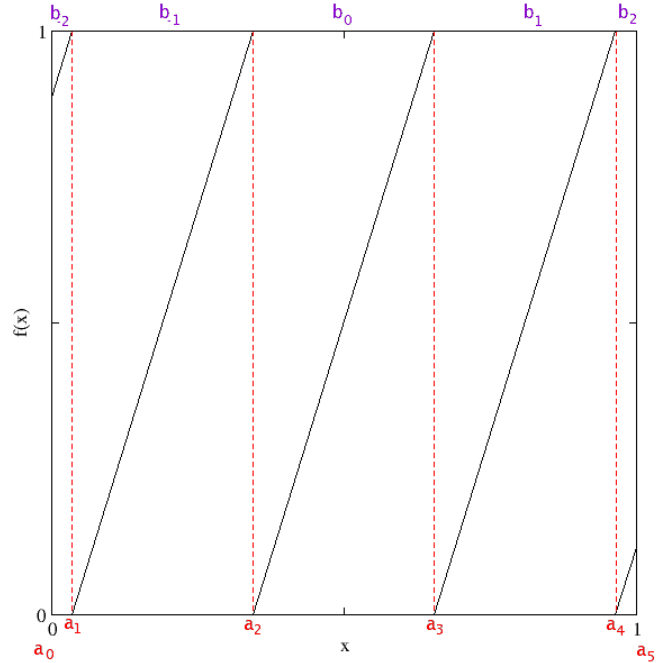}}
\vglue -0.3cm 
\caption{Example of the map $f_{\Lambda}(x)$ with $\Lambda=3.23$.}
\label{fig1}       
\end{figure}

A bounce of initial condition for the map $g(x)$ will spread out diffusively on the real line and the set of transport 
coefficients (for example the drift or the diffusion coefficient) can be computed 
with the help of a generalized  dynamical zeta function \cite{POE} of the form:
\begin{equation}\label{zeta}
\zeta^{-1}(z,\beta)=\prod_{\{ p \} }\left(1-\frac{z^{n_p}e^{\beta \sigma_p}}{|\Lambda_p|}\right)
\end{equation}
where  the product runs over prime periodic orbit of the reduced map $f(x)$ and the only quantities that
enter the definition are the period of the orbit $n_p$, the instability $\Lambda_p=\prod_{i=0}^{n_p-1} f'(f^i(x))$ and the integer jumping number $\sigma_p=g(x)-f(x)$ (that remembers the number of cells the trajectory  jumps once unfolded on the real line).  Cycle expansions technique \cite{POE}  provides an organization of the infinite set of periodic orbits $\{p\}$ that produces (for fully hyperbolic systems) a rapidly convergent power expansion of the dynamical zeta function (\ref{zeta}). 
Transport properties of the system are  encoded into  the behaviour of its smallest zero $z_0(\beta)$ around its value $z_0(0)=1$. For example the drift J and the diffusion coefficient D can be computed by
\begin{equation}\label{transport-c}
J=\left.\frac{{\rm d} z_0(\beta)}{{\rm d} \beta} \right|_{\beta=0} \qquad D=\frac12 \left.\frac{{\rm d}^2 z_0(\beta)}{{\rm d} \beta^2} \right|_{\beta=0}
\end{equation} 

 For odd integer $\Lambda$  the map consists of  complete branches and the symbolic dynamic reduces to a full shift on an $n$-letter alphabet. In such a case the dynamical zeta function can be written in closed form and the diffusion coefficient easily computed. The situation is similar for even integer. For a specific set  of the parameter  $\Lambda$ it was possible \cite{japan}   to control the symbolic dynamic and construct the dynamical zeta function while for the generic case we will expect an 
infinite number of pruning rules (changing discontinuously under parameter variation)  and no simple expression for the cycle expansions organization of periodic orbits.  On the other hand  
Milnor and Thurston \cite{MiThu}  have shown how it is possible to control the symbolic dynamic of  maps of the unit interval following the iteration of critical points. In particular they were able to  relate the topological entropy to the determinant of a finite matrix (the kneading determinant) were the entries are formal power series (with coefficients determined by the kneading trajectories). Later  Baladi and Ruelle \cite{BaRu} have generalized the result to constant weight (see also \cite{b-etal} and references therein for more general results). We will closely follow the notation of Baladi-Ruelle \cite{BaRu} to construct the dynamical zeta function for  the family of  maps $f_{\Lambda}(x)$ defined above.  
 
 Fix $\Lambda$ and rename for simplicity $f(x):=f_{\Lambda}(x)$.  Call  $a_0<a_1<...<a_N$  the ordered sequence of end points of each branch (see Fig.(\ref{fig1})). Define $\epsilon(x)= \pm1$, whether $f(x)$  is increasing or decreasing and $t(x)$ as a constant weight  for $x \in [a_{i-1},a_{i}]$. In our specific case the restriction of these functions on the $i$-interval will take the constant values:
 \begin{eqnarray}\label{weights}
\epsilon_i&=&1 \\
 t_i&=&\frac{z \,e^{\beta \sigma_i}}{\Lambda}  
 \end{eqnarray} 
 where $\sigma_i$ is the jumping number associated to the i-th branch.
 
 Let's associate to each point $x$ the address vector in $\mathbb{Z}^{N-1}$
 \begin{equation}\label{address}
 \vec{\alpha}(x)=\left[{sgn}(x-a_1),...,{sgn}(x-a_{N-1})\right]
 \end{equation}
 and define the invariant coordinate of $x$ by the formal series:
 \begin{equation}
 \vec{\theta}(x)=\sum_{n=0}^{\infty}\left[\prod_{k=0}^{n-1}(\epsilon t)(f^k(x))\right]  \vec{\alpha}(f^n(x))
 \end{equation}
 where the product is intended equal to one if  $n=0$ (the invariant coordinate $\theta$ is single valued once we put $\epsilon(a_i)=0$).
 
Defining $\phi(a^{\pm}) =\lim_{x \to a^{\pm}} \phi(x)$ we compute the discontinuity vector at the critical points $a_i$ for $i=1,..,N-1$:
\begin{equation}
\vec{K}_i(z,\beta) =\frac12\left[\vec{\theta}(a_i^+)-\vec{\theta}(a_i^-)\right]  
\end{equation}
The  kneading matrix $K(z,\beta)$ is defined as the $(N-1) \times (N-1)$ matrix with $\vec{K}_i; \,i=1,.., N-1$ as rows.
Let's call 
$\Delta(z,\beta)=\det{K(z,\beta)}$ 
the kneading determinant.

\section{ Relation between kneading determinant and dynamical zeta function}
Without entering into many details we restrict to the case under study and present the formula that relates the kneading determinant and the dynamical zeta function for the map $f(x)$ defined above.  Let' s call $\{\tilde{p}\}$ the set of prime periodic orbits that have a critical point as periodic point (i.e. $f^{n_{\tilde{p}}}(a_i)=a_i$). 
 We note explicitly that there is at most a finite number of such prime orbits (for finite number of $a_i$).
 It is possible to show that the dynamical zeta function is equal to the kneading determinant up to a rational function (see \cite{BaRu} for more general results and a proof of relation (\ref{magic})):

\begin{eqnarray}\label{magic}
\Delta(z,\beta)&=&R(z,\beta)\, \zeta^{-1}(z,\beta)\\
R(z,\beta)&=&\left[1-\frac12(\epsilon_1t_1+\epsilon_Nt_N)\right]\prod_{  \{\tilde{p} \} }[1-t_{\tilde{p}}(z,\beta)]^{-1}
\end{eqnarray}

In their original work Baladi and Ruelle inserted the product over the set  $\{\tilde{p}\}$ into a definition of a reduced zeta function where they do not count such (repelling) cycles. Here we prefer to remain with the usual zeta function and factorize out explicitly the critical cycles. In particular formula (\ref{magic}) shows that the smallest zero of the kneading determinant coincides with the smallest zero of the dynamical zeta function  for $\beta \to 0$.

\section{Fractal diffusion coefficient}

Let's start computing explicitly the kneading determinant for the map $f_{\Lambda}$.
It's easy to see that, for $i=1,..,N-1$:
\begin{eqnarray}
\vec{\theta}(a_i^+)&=&(....,+,+,-,....)+t_{i+1}\vec{\theta}(a_0^+)\\
\vec{\theta}(a_i^-) &=&(....,+,-,-,....) + t_{i}\,\vec{\theta}(a_N^-)
\end{eqnarray}
where the change of sign in the component of the first addend appear at the i-th position.
From the choice (\ref{weights}) we've $t_{i+1}=e^{\beta}t_i$ and so
\begin{equation}
\vec{K}_i(z,\beta)=(0,...,0,1,0,....,0)+\frac12 t_i e^{\frac{\beta}2} \left[e^{\frac{\beta}2} \vec{\theta}(a_0^+)-e^{-\frac{\beta}2} \vec{\theta}(a_N^-)\right]
\end{equation}
Simplifying the notation we get that the kneading matrix is of the form:
\begin{equation}
K_{ij}=1-W_{ij}  \quad \quad W_{ij}=a_ib_j\label{W}
\end{equation}
and the kneading determinant  can be easily computed to be
\begin{eqnarray}
\Delta&=&\det(K)=\exp{\left[-\sum_{n=1}^{\infty} \frac{ {\rm Tr}\, (W^n)}n\right]}=\\
&=&1-{\rm Tr}\, W
\end{eqnarray}
where in the last passage we've used the fact that for  matrix W of the form (\ref{W}) we've	
 ${\rm Tr}\, (W^n)= ({\rm Tr}\, W)^n$. 
Finally the kneading determinant simplify to:

\begin{equation}
\Delta(z,\beta)=1+\frac{z}{2\Lambda}\sum_{i=1}^{N-1}{e^{\beta(\sigma_i+1/2)}\left[e^{\frac{\beta}2} \vec{\theta}(a_0^+)-e^{-\frac{\beta}2} \vec{\theta}(a_N^-)\right]_i}
\end{equation}

Due to the symmetry under $x \to -x$ the map $g(x)$  presents no net drift. In order to show it we   compute  the first derivative of the smallest zero of the kneading determinant (\ref{transport-c}).  Let's relabel the intervals by the associated jumping numbers:  call it $b_j \quad j=-M,...,0,..,M$ and call  $\vec{\alpha}(j)$ the address (\ref{address}) of a point in the interval $b_j$. Let's define the function $s_j(z,\beta)$ by the relation:
\begin{equation}
e^{\beta/2}\vec{\theta}(a_0^+)=:\sum_{j=-M}^M s_j(z,\beta) \vec{\alpha}(j)
\end{equation}
In the symmetric case  iterations of the critical point $a_0$ can be mapped to iterations of the other 'external' critical point $a_N$ and so 
\begin{equation}
e^{-\beta/2}\vec{\theta}(a_N^-)=\sum_{j=-M}^Ms_j(z,-\beta) \vec{\alpha}(-j)
\end{equation}
With the definition above  the function $s_j(z,\beta)$  receives contributions only  when the trajectory of the critical point $a_0$ traverses the interval $b_j$. 

Each of the $\vec{\alpha}(j)$ has $2M$ components (see \ref{address}) that, with abuse of notation, we can label by $k=[(-M+\frac12),...,(-2+\frac12),-\frac12; \frac12, (2-\frac12),..,(M-\frac12)]$. The kneading determinant  can be rewritten as:

\begin{equation}
\Delta(z,\beta)=1+\frac{z}{2\Lambda}\sum_{k}\sum_{j} e^{\beta k}\left[
s_j(z,\beta) \vec{\alpha}(j)-s_j(z,-\beta) \vec{\alpha}(-j)\right]_k
\end{equation}
and the first derivative is then
\begin{eqnarray}\label{first}
\left.\frac{\partial\Delta(z,\beta)}{\partial{\beta}}\right|_{z=1,\beta=0}&=&\frac{z}{2\Lambda}\left.\sum_{k,j} \Big[s_j(1,0)(k+\frac12)  \vec{x}^j_k+\partial_{\beta}s_j(1,0)\vec{y}^j_k\Big]\right|_{z=1,\beta=0}\\
&=&0
\end{eqnarray}
were we've used:
\begin{eqnarray}
\vec{x}(j)&=& [\vec{\alpha}(j)-\vec{\alpha}(-j)] ; \qquad \vec{y}(j)= [\vec{\alpha}(j)+\vec{\alpha}(-j)]
\end{eqnarray}
\begin{eqnarray}
[\vec{x}(j)]_{k} &=&[\vec{x}(j)]_{-k}  ;  \qquad \qquad      [\vec{y}(j)]_k=-[\vec{y}(j)]_{-k}
\end{eqnarray}
 
For $\beta=0$ the leading zero of the dynamical zeta function must be  $z(0)=1$ and then  the kneading determinant must satisfy: 
\begin{eqnarray}
\Delta(1,0)&=&1+\frac{1}{2\Lambda} \sum_{k,j}
s_j(1,0) [\vec{x}(j)]_k=\\
&=&1+\frac{2}{\Lambda} \sum_{j=-M}^M j\,s_j(1,0)=0\label{uno} 
\end{eqnarray}
where in the last passage we've use the fact that $\sum_k{[\vec{x}(j)]_{k}}=4j$.
In order to clarify Eq. (\ref{uno})  let's check it for the simplest case of odd integer $\Lambda$. In such a case the map consists of complete branches and $a_0=0$ become a fixed point. The function $s_j(z,\beta)$ vanishes for all indexes   but $j=-M$ where
 \begin{equation}
s_{-M}(z,\beta)=\sum_{n=0}^{\infty}\left[\frac{z \, e^{-\beta M}}{\Lambda}\right]^n \, e^{-\frac{\beta}2}
\end{equation}
Using the fact that $M=(\Lambda-1)/2$ the relation (\ref{uno}) becomes
\begin{eqnarray}
\Delta(1,0)&=&1-\frac{1}{\Lambda}(\Lambda-1)\frac{\Lambda}{\Lambda-1} =0
\end{eqnarray}

Similar relations are found for even integer $\Lambda$ were again the address of $a_0$ can be written in closed form due to the fact that $f^n(0)=1/2 \quad \forall n>0$.

\begin{figure}[h!]  
\centerline{\epsfxsize=13cm\epsffile{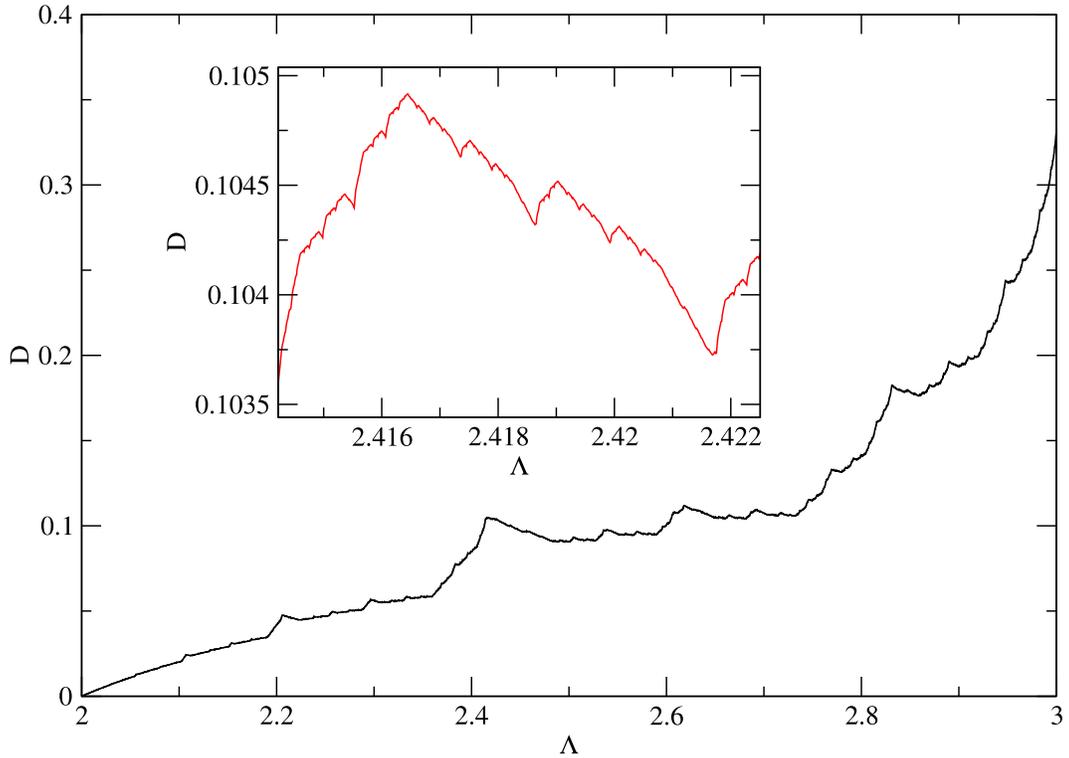}}
\vglue -0.3cm 
\caption{Fractal diffusion coefficient as a function of the slope of the map $f_{\Lambda}(x)$ computed from the smallest zero of the dynamical zeta function. The insert is a blow up of a part of the main figure}
\label{Dnum}       
\end{figure}

In the same spirit it is possible to compute the diffusion coefficient for a generic value of $\Lambda$, rewriting Eq. (\ref{transport-c}) as
\begin{equation}
D=-\frac12\left.\left(\frac{\partial^2\Delta(z,\beta)}{\partial{\beta^2}}/\frac{\partial \Delta(z,\beta)}{\partial z}\right)\right|_{z=1,\beta=0}\label{D}
\end{equation}
where
\begin{eqnarray}
\left.\frac{\partial^2\Delta(z,\beta)}{\partial{\beta^2}}\right|_{1,0}&=&\frac{1}{2\Lambda}\sum_{k,j} \Big[k^2s_j(z,\beta)\vec{x}(j)_k+2k\frac{ \partial s_j(z,\beta)}{\partial \beta}\vec{y}(j)_k+\nonumber\\
&&\left.\qquad \qquad \qquad \qquad \qquad   +\frac{\partial^2 s_j(z,\beta)}{\partial \beta^2} \vec{x}(j)_k\Big]\right|_{1,0}\\
\left.\frac{\partial \Delta(z,\beta)}{\partial z}\right|_{1,0}&=&-1+\frac1{2\Lambda}\left.\sum_{k,j}{ \frac{\partial s_j(z,\beta)}{\partial z} \vec{x}(j)_k}\right|_{1,0}
\end{eqnarray}
The evaluation of expression (\ref{D}) for a parameter choice $\Lambda \in [2,3]$ is shown in Fig. (\ref{Dnum}) and can be compared, for example, with Fig.10 in \cite{Klages-Dorfman} . 
\vspace{1.1cm}

\section{Conclusions}
The fractal diffusion coefficient that  appears in some simple piecewise linear map is the consequence of the sensitivity of the topological entropy to parameter variation: even arbitrary small parameter change  creates and destroy infinitely many periodic orbits and the cycle expansions approach to dynamical zeta function  becomes hard to follow. Avoiding  periodic orbits and using the Baladi-Ruelle generalization of the 
Milnor-Thurston kneading determinant we were able to construct the dynamical zeta function for arbitrary parameter value and compute from it the fractal diffusion coefficient.

\ack
I would like to thank R. Artuso for helpful comments and suggestions and R. Klages for interesting discussions.



\section*{References}
\begin{thebibliography}{99}

\bibitem{POE} Artuso R, Aurell  E and  Cvitanovi\'c P 1990 {\it Nonlinearity }{\bf 3} 325;
Artuso R,  Aurell E and Cvitanovi\'c P 1990 {\it Nonlinearity} {\bf 3} 361;
Cvitanovi\'c P, Artuso R,  Dahlqvist P,  Mainieri R, Tanner G,  Vattay G, Whelan N and  Wirzba A 2003 {\it Chaos: classical and quantum} ({\tt www.nbi.dk/ChaosBook/} )

\bibitem{inter} Prellberg T  2003 {\em J. Phys. A: Math. Gen \bf 36} 2455-2461;
Isola S  2002 {\em Nonlinearity \bf 15} 1521-1539; 
Artuso R, Cvitanovi\'c P and Tanner G 2003 {\em Prog.  Theor. Phys. Suppl. \bf 150} 1

\bibitem{vari} Artuso R, Casati G and Lombardi R  1993  {\em Phys.Rev.Lett. \bf 71} 62-64;
Dahlqvist P and Artuso R 1996 {\em Phys.Lett.A \bf 219}  212;
Dahlqvist P 1999 {\em Phys. Rev. E \bf 60} 6639-6644;
Artuso R and Cristadoro G  2004  {\em  J. Phys. A: Math. Gen \bf 37} 85-103 

\bibitem{Klages-Dorfman}{Klages R and Dorfman J R 1999 {\em Phys. Rev. E \bf 59}  5361Ð5383 
}

\bibitem{Gro}{Groeneveld J and Klages R  2002 {\em J. Stat. Phys. \bf 109}  821-861} 

\bibitem{japan} Tseng H-C, Chen H-J, Li P-C, Lai W-Y, Chou C-H and Chen H-W 1994 {\em Phys. Lett. A \bf 195}, 74-80;
Chen C-C 1995  {\em Phys. Rev E \bf 51} 2815-2822;
Tseng H-C and Chen H-J Chen 1996 {\em Int. J. Mod. Phys. B \bf 10} 1913-1934 



\bibitem{MiThu}{Milnor J and Thurston W 1988 {\it Lectures Notes in Math. \bf 1342}, 465-563;
}
 
\bibitem{BaRu}{Baladi V and Ruelle D 1994 {\em Ergodic Theory Dynamical Systems \bf 14} 621-632  
}

\bibitem{b-etal}{ Baladi V and Ruelle D 1996 {\em Invent. Math. \bf 123} 553-574 
Baillif M and Baladi V 2004  [{\it Preprint} math.DS/0211343] 
}

\end{thebibliography}
\end{document}